# BIRD'S-EYE VIEW ON NOISE-BASED LOGIC

LASZLO B. KISH [1], CLAES G. GRANQVIST [2], TAMAS HORVATH [3], ANDREAS KLAPPENECKER [4], HE WEN [5], SERGEY M. BEZRUKOV [6]

[1] *Department of Electrical and Computer Engineering, Texas A&M University, College Station, Texas 77843-2128, USA; email: laszlo.kish@ece.tamu.edu*

[2] *Department of Engineering Sciences, The Ångström Laboratory, Uppsala University, P.O. Box 534, SE-75121 Uppsala, Sweden; email: claes-goran.granqvist@angstrom.uu.se*

[3] *Department of Computer Science, University of Bonn, Germany; Fraunhofer IAIS, Schloss Birlinghoven, D-53754 Sankt Augustin, Germany; email: tamas.horvath@iais.fraunhofer.de*

[4] *Department of Computer Science, Texas A&M University, College Station, Texas 77843-2128, USA; email: klappi@cse.tamu.edu*

[5] *College of Electrical and Information Engineering, Hunan University, Changsha, China. email: he_wen82@126.com*

[6] *Laboratory of Physical and Structural Biology, Program in Physical Biology, NICHD, National Institutes of Health, Bethesda, MD 20892, USA*

Noise-based logic is a practically deterministic logic scheme inspired by the randomness of neural spikes and uses a system of uncorrelated stochastic processes and their superposition to represent the logic state. We briefly discuss various questions such as (*i*) What does practical determinism mean? (*ii*) Is noise-based logic a Turing machine? (*iii*) Is there hope to beat (the dreams of) quantum computation by a classical physical noise-based processor, and what are the minimum hardware requirements for that? Finally, (*iv*) we address the problem of random number generators and show that the common belief that quantum number generators are superior to classical (thermal) noise-based generators is nothing but a myth.

Noise-based logic (NBL) [1–13] is a practically deterministic logic scheme inspired by the randomness of neural spikes and uses a system of uncorrelated stochastic processes and their superposition to represent the logic state. "Practically deterministic" means that the results emerge with non-zero error probability, but this error probability decays exponentially with increasing observation time. In this short Note we briefly summarize a few key aspects of NBL.

## 1. Justifications to explore noise based logic

### (*a*) *Energy dissipation*

NBL has a potential to reduce power dissipation of logic operations [1,13], which is a consequence of Brillouin's negentropy principle [13]. Such dissipation is unavoidable in information systems [14,15] and exists even in neural systems [16], but today's computers use many orders-of-magnitude more power than the one given by the fundamental limit of $kT \ln\left(\frac{1}{\varepsilon}\right)$ per bit energy dissipation [14,15], where $\varepsilon \ll 0.5$ is the error probability of bit operation.

(**b**) *Exponentially large logic depth and exponential speed-up*

For some special-purpose operations, properly designed NBL engines provide not only exponential logic depth [2,8,9,13] but also exponential speed-up in instantaneous logic systems [8,9,13].

(**c**) *NBL is a Turing computer with ideal random number generation*

The strong Church-Turing Theorem (SCTT) states [17] that

(*i*) any "reasonable" model of computation can be efficiently simulated on a probabilistic Turing machine, and

(*ii*) no computer can be more efficient than a digital one equipped with a random number generator.

Here the definition of relative efficiency is that [17]

(*iii*) computer A is "more efficient" than computer B if A can solve, in polynomial time, a problem that cannot be solved in polynomial time by computer B.

Whereas the creation of NBL was not inspired by the SCTT but by the stochastic neural signal components of the brain, the SCTT is relevant because

(*iv*) discrete-amplitude versions of NBL, including the instantaneous NBL and brain-logic schemes, can be realized by Turing machines (digital computers) equipped with one or more random number generators.

(*d*) *The brain is a biological representation of NBL*

The logic signals in the brain are stochastic, which has inspired various NBL-based representations and raised many relevant questions [3–5]. NBL-based string verification schemes, generalized for the brain, show how intelligence leads to reasonable decisions based on a very limited amount of information [5]. These results provide a conceptual explanation of the reason why spike transfer via neurons usually is statistical with less than a 100% success rate.

## 2. Can NBL realize some of the quantum computing dreams, or more?

NBL has already realized a number of quantum computing dreams. Thus

(*v*) no decoherence problems of any kind are present, and hence error correction is not needed,

(*vi*) superposition of $2^N$ integer numbers can be accomplished with a simple operation [2] containing only about $2N$ algebraic operations, and

(*vii*) exponentially fast bit-operations can be executed instantaneously on the superposition of $2^N$ classical bits with low $O(2N)$ hardware and time complexity [8,9,13]; all of the single-bit quantum gates have successfully been implemented and have this factor of $2^N$ exponential speed-up [13].

(*viii*) To realize a NBL processor that is equivalent to a quantum computer with 200 effective qubits, only simple hardware is needed [8]: a classical binary computer (Turing machine) with an algorithm that can handle 400 bits accuracy and a physical random number generator. Error correction is not required. These involve relatively small efforts/costs; the real effort is needed for the development of useful special-purpose algorithms.

## 3. How about random number generation? Classical thermodynamic, or quantum?

There is a common belief, expressed for example by Frauchiger *et al*. [18], that quantum physics is needed for random-number generators to be "really random". The root of this belief seems to be a notion that quantum randomness is "inherent" or can be "proven", whereas classical physics is deterministic. It should be noted that the second claim implies that thermal noise is not random if one can access the initial (and boundary) conditions of the elements of the system.

The above-mentioned belief is fundamentally flawed, as explained next:

(*ix*) Quantum randomness cannot be proven within quantum physics but is an axiom; the Born-interpretation of the wave function, as well as axioms, can never be scientifically proven.

(*x*) Classical physics is indeed deterministic, but thermal (Johnson) noise in a pure conductor crystal is not. Instead this noise is due to the random motion of electrons, which is disrupted by acoustical phonon scattering [19] that has no phase memory because it is inelastic. This means that, even if somebody is able to determine the initial conditions for all of the $\sim 10^{20}$ electrons as well as the initial conditions of all oscillatory lattice modes in the sample, this information disappears within the mean-free time of electron transport, which is of the order of $10^{-13}$ seconds in conductors. Furthermore, all of the information is totally erased during this time as a consequence of the axiom referred to above—*i.e.*, Born's interpretation of quantum physics—since lattice scattering is a quantum phenomenon. Thus the fundamental randomness of Johnson noise has the same foundation as the one underlying quantum random number generators, though with a great added benefit: the extraordinarily large number of degrees of freedom by non-linear mixing due to collision processes. Similar arguments hold for defect scattering in imperfect and/or dirty crystals, except that the loss of information takes a number (about 100) of scattering events [19], which still results in a very short time ($\sim 10^{-11}$ seconds) at room temperature.

(*xi*) Quantum number generators, such as polarization beam splitters with two photodiodes, have only two degrees of freedom and are extremely vulnerable to (*a*) mechanical vibrations, which introduce a long periodic bias, (*b*) laser fluctuations of polarization, intensity profile, *etc*, for which the dominant components are various $1/f$ noise processes with logarithmically decaying correlation function and providing strong memory for long times, and (*c*) all of the similar $1/f$-type noises with long memory in detectors and preamplifiers.

In conclusion, quantum random-number generators are poor concepts. They would compete with thermal noise if $\sim 10^{20}$ independent systems could be integrated on a chip to supply a single random-number series instead of employing the usual single bulky system. However, such integration is impossible due to the large wavelength of photons, and for practical reasons. At the same time, random-number generation exists naturally within the processes underlying Johnson noise.

Finally, how can one create good physical random-number generators:

(*xii*) The simplest solution is to integrate a large number of thermal noise-based random-number generators on a chip. Such a generator can be, for example, the XOR function for the sign of the amplitude and velocity of Johnson noise. Due to their Gaussianity, the amplitude and velocity are independent processes, and even a carefully designed single generator of this kind can be satisfactory and pass all available randomness tests. The use of a large number of generators, and implementing the XOR function to multiply their bit output, results in a classical-physics-based random-number generator that is superior to any known solution.

**Acknowledgements**